\documentclass[12pt]{iopart}
\usepackage{graphicx,graphics}

\begin{document}

\title[First steps to Virtual Mammography]{First steps to Virtual Mammography: Simulating external compressions of the breast with the Surface Evolver}

\author{M Z Nascimento$^1$ and V Ramos Batista$^2$\footnote{This work is dedicated to my wife Clarice.}}

\address{
$^1$ FACOM-UFU, Av. Jo\~ao Naves de \'Avila 2121, Bl.A, 38400-902 Uberl\^andia-MG, Brazil. E-mail: nascimento@facom.ufu.br\\ 
$^2$ CMCC-UFABC, R. Sta. Ad\'elia 166, Bl.B, 09210-170 St. Andr\'e-SP, Brazil. E-mail: valerio.batista@ufabc.edu.br}

\begin{abstract}
In this paper we introduce a computational modelling that reproduces the breast compression processes used to obtain the mammogram. The main result is a programme in which one can track the first steps of virtual mammography. On the one hand, our modelling enables addition of structures that represent different tissues, muscles and glands in the breast. On the other hand, we shall validate and implement it by means of laboratory tests with phantoms. To the best of our knowledge, these two characteristics do confer originality to our research. This is because their interrelation seems not to be properly established elsewhere yet. We conclude that our model reproduces the same shapes and measurements really taken from the volunteer's breasts.
\end{abstract}

\vspace{2pc}
\maketitle

\section{Introduction}

The obit rate caused by cancer around the globe is tracked by the World Health Organization. Their statistics indicate that 7.6 million of the obits in 2008 were due to cancer. This ratio has been increasing and some estimates show that it can reach 13.1 million cases until 2030. Lung cancer in women is still the leading obituary kind and breast cancer follows right next. See {\tt http://www.who.int/mediacentre/factsheets/fs297/en} \ for details.

Medical imaging methods help detect breast cancer. The most used are resonance, ultrasonography and mammography. However, screen/film mammography is the primary imaging modality due to low cost and accessibility. It has proved to be an effective aid in the early detection and most of the restraint in obit rate is solely thanks to this method. But it shares a disadvantage with the others: in the imaging procedure forces applied to the breast change its shape considerably. For the surgery tumour location is highly uncertain, and so a great portion of the breast has to be removed \cite{P}.

Along the past decade there has been an effort to develop softwares devoted to anatomic modelling of the breast and of deformations by mammography procedures. Among others we cite \cite{B,H,Ku,N,P} and \cite{T}. Some are aesthetically impressive, like \cite{BS}. For a model at rest this latter is indeed very complete, but its functionality has not been presented yet. The problem will reside in the excessive computational complexity at managing all different elements that take part in the model. The source codes tend to be untreatable in such cases.

In order to bypass these difficulties perhaps a simpler approach like the ITK software of \cite{P} turns out to be more accessible. However, the human breast is composed by different tissues, muscles and glands, as highly taken into account in \cite{BS}. Since ITK works with three dimensional meshes, it is a hard task to separate all those components into specific clusters. Moreover, this produces a ``Lego effect'' that demands too much orchestration to control the belonging of each mesh element to each tissue, muscle or gland.

Figures 1 and 2 show how problematic this approach is. For instance, consider their representation of the mammal glands. They are numerous and independent. Softwares like ITK treat them as a sole component, or sometimes even omit them by considering that the fat tissue composes the whole interior. 

\centerline{
\includegraphics[scale=0.70]{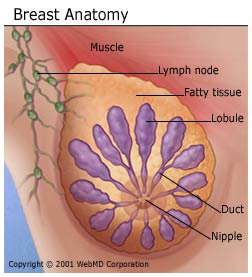}\hfill
\includegraphics[scale=0.08]{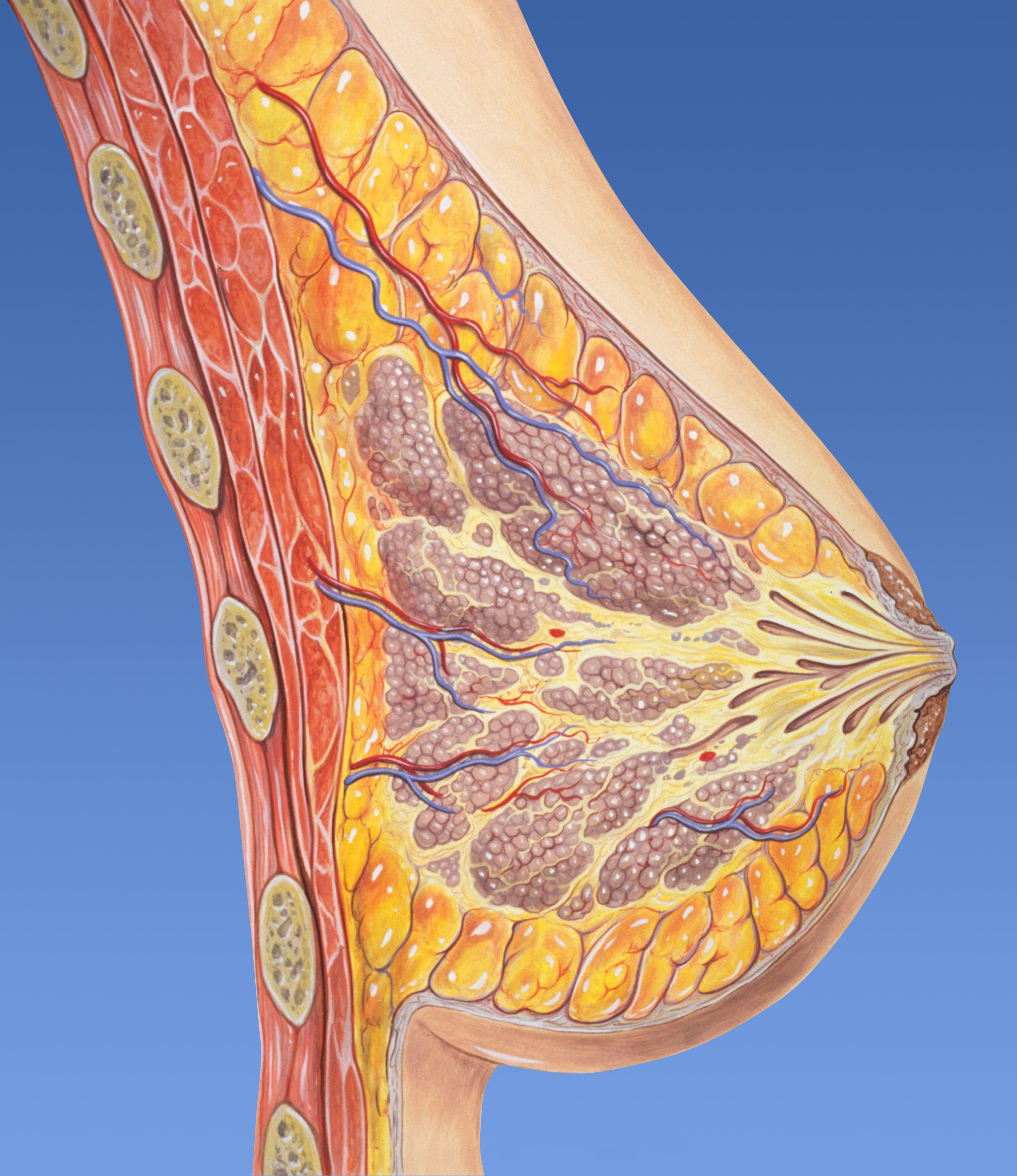}}

\centerline{Figure 1: Front view of the breast\footnote{Taken from {\tt http://www.webmd.com/breast-cancer/guide/breast-cancer-normal-breast}}.\hfill Figure 2: Side view\footnote{Taken from {\tt http://en.wikipedia.org/wiki/Portal:Medicine/Selected\_picture/26}}.}
\ \smallskip

Which simplifications can be made without going into considerable errors? Regarding numeric modelling of the breast deformation during mammography, we still do not have enough satisfactory answers. In this case, the dilemma between complexity and simplification motivated us to work with the {\it Surface Evolver} \cite{B1}. Its computational representation of any element is made by surface {\it layers} that enclose it. Evolver allows to add as many elements as needed, together with their individual attributes like mass, geometric restrictions, contact and elastic bending energies, etc. Figure 3 shows one of our initial Evolver models, when the patient lies horizontally and her breast is supported by a surgical drape. We call it the {\it surgery} position. Figure 4 shows the deformation resulted from gravity, thorax curvature, breast weight and dimensions when the patient stands up with her spine straight. We call it the {\it standing-up} position. This picture was enhanced with {\it Geomview}. See {\tt http://www.geomview.org} for details. All data were obtained from the volunteer.

\includegraphics[scale=0.25]{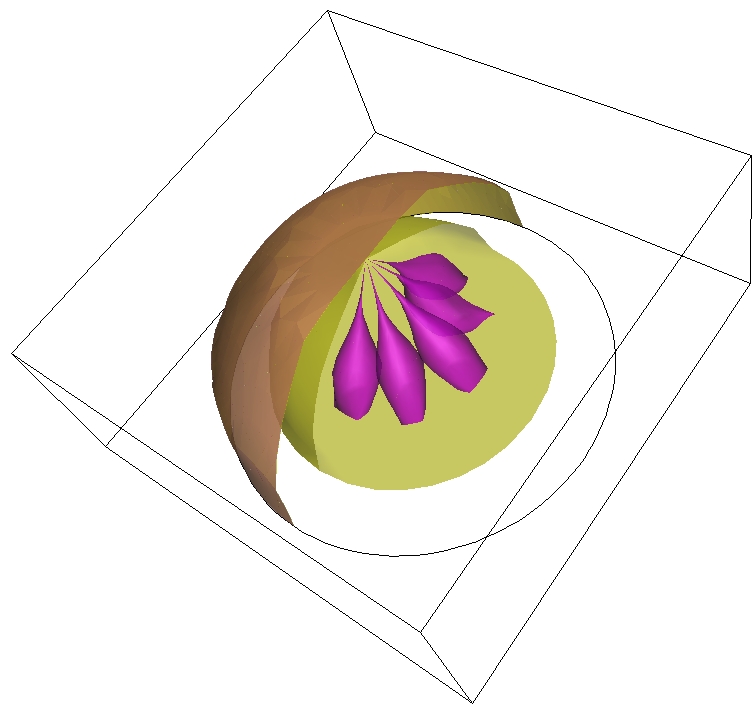}\hfill
\includegraphics[scale=0.34]{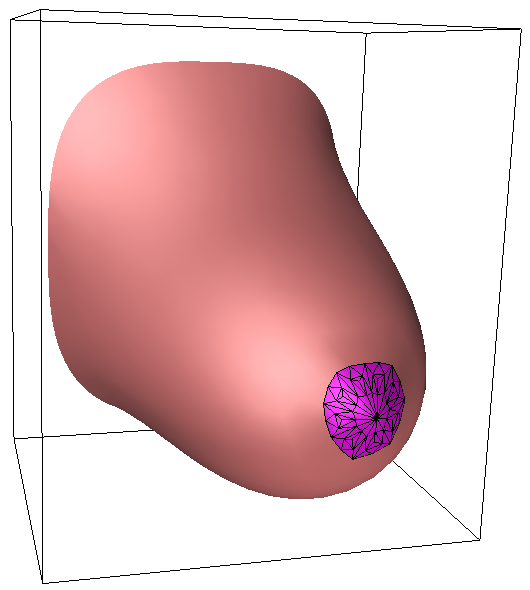}

\centerline{Figure 3: Surgery position with visible elements.\hfill Figure 4: Standing-up position.}
\ \smallskip

The {\it Surface Evolver} is largely applied in several Areas of Knowledge, like Biology, Physics, Chemistry and Mathematics. For instance, see \cite{C}, \cite{GB} and \cite{S}. It enables to control the triangulation of each surface layer at any step of the simulation and plainly. This way we can make triangles uniform, refine and remove them either locally or globally, in order to avoid lack and excess of graphical data, as happens to some pictures from \cite{P}. Evolver also enables to include optimization of quantities given by integrals, as well as to track values like of pressure, area and volume at any step of the simulation. 

However, it is important to remark that Evolver works with {\it surfaces}. Even when simulating volume, mass, gravity, etc., the inside of surfaces are meant to be either {\it ideal gases} or {\it liquids}. Namely, it was not programmed for three dimensional elasticities, but for superficial elasticities.

In the long-term our research aims at a full and detailed reproduction of the mammography procedure with Evolver. By starting with the mammographies that show tumour nodules, we rewind the virtual procedure. This way nodules will be located with precision for the surgery.

For the time being, our work is devoted to simulations of external compressions of the breast. In order to track tumour nodules we shall either consider the complexity of internal parts, or equate their trajectory by performing experiments with transparent breast phantoms. Evolver allows to increase complexity up to three dimensional meshes, but if this turns out to be intractable or inefficient, we shall adopt the numerical trajectories and pre-define them in our Evolver model. See Subsection 2.2 for details.

Anyway, no matter how accurately we can locate a nodule the surgeon will always remove some tissue that surrounds it. This prevents from {\it infiltration}, which is an abnormal accumulation of cells around the cancer. 

The readers will be able to run some of our datafiles with Evolver. In the Appendix you will find all details about accessing this software through our virtual machine.

\section{Methods and materials}

\subsection{Taking measurements from a real patient}

In order to take mammographies in our virtual environment we have subdivided the whole process in 6 main steps for each breast. We call them SRG ({\it surgery}), STU ({\it stand-up}), LAT ({\it lay-on-table}), CRC ({\it cranio-caudal}), LET ({\it lean-on-table}) and MLO ({\it medio-lateral-oblique}). 

Of course, at each step the mass of the breast remains constant while its shape changes drastically. Except for CRC and MLO, changes in the volume and average density are much slighter than in the area. There is enough literature with average values, but for our purposes they are inadequate because measurements vary a lot from woman to woman. 

It is possible to apply electrodes to draw a computer model and then get magnitudes numerically. However, this procedure is not practical and requires expensive equipment.

But we check our data in Evolver by means of a virtual tape-measure that we call TMR. Our procedure is to enter the initial values and then use the TMR to calibrate them. For the time being, the calibration is still interactive. We shall make it automatic in future.

Figure 5 left shows a standing woman with her spine straight. The dotted line indicates a tape-measure that passes {\it under} her breast. With the open right hand upon her belly we locate our coordinates: $Ox$ points as the thumb, $Oy$ points as the other fingers, and the palm is raised along $Oz$. Figure 5 right shows the oval curve obtained after having taken the volunteer's measurements. Important values are indicated there, and we have fixed all magnitudes in the cgs-system. Notice that the overall perimeter is 86cm.

\centerline{
\includegraphics[scale=1.50]{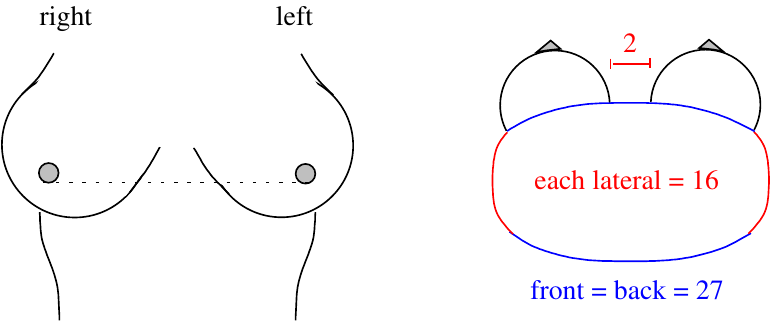}}

\centerline{Figure 5: Measuring strategic stretches of the thorax.}
\ \smallskip 

Some other important values are listed in Figure 6. The CRC and MLO measurements are taken directly from the mammographies.

In Figure 6 LAT is identified by the contour of a typical transparent table used in mammography. Its dimensions are standard: 18 by 24 centimetres. Notice that all horizontal and vertical arcs are given by a sum of two numbers. These correspond to semi-arcs, from the left to the right horizontally and from bottom to top vertically. The semi-arcs meet at the nipple and can be used to check if our simulator locates it correctly.

Our simulation with Evolver always start at SRG because it is the most symmetric position. In practice, not only a surgical drape holds the base of the breast but also an auxiliary surgeon and sometimes an extra support fitted under the woman's armpit.

By the way, the measurements depicted in Figure 6 are {\it meant to be taken with this extra support}, which can be a wrapped hand towel. As an example, the breast perimeter of the volunteer varied from 46cm to 50cm when the support was removed. This is due to the fact that {\it the tape-measure can never be tightened around the parts}. 

Another position that is very likely to be symmetric is LAT. However, before taking the X-ray snapshot sometimes the woman is asked to slightly bend her knees and/or twist her back, but still keeping her spine straight. Since all these variations result in too small changes in the measurements, we have opted for the most symmetric LAT.
\\

\includegraphics[scale=1.50]{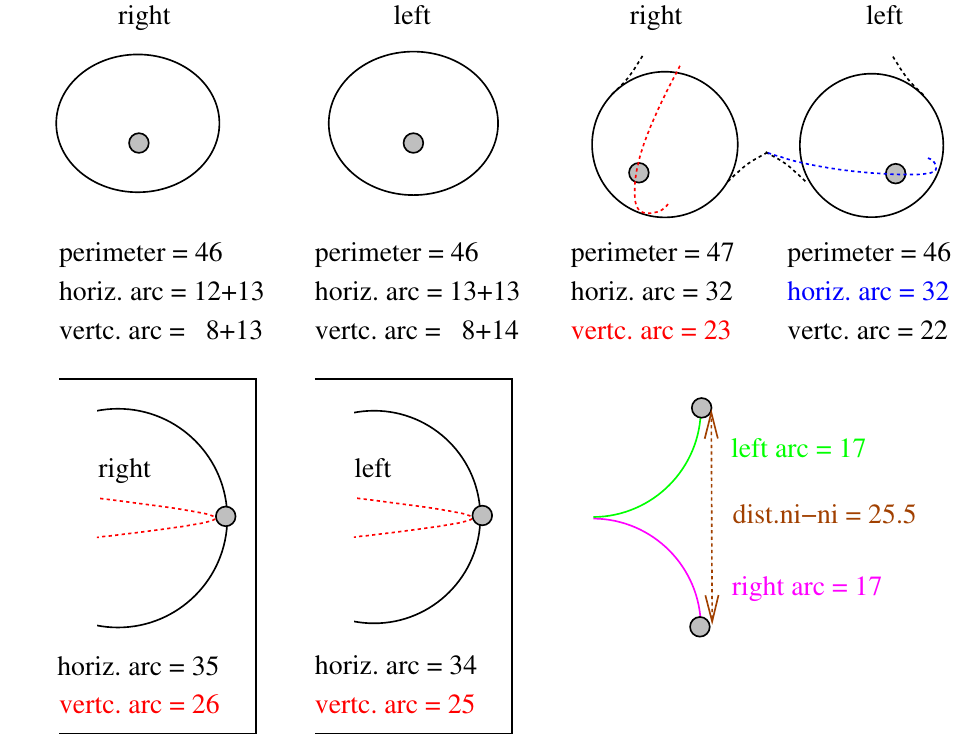}

\centerline{Figure 6: Measurements from SRG, LAT and STU, this latter in colours.}
\ \smallskip

Our Evolver datafiles follow the sequence SRG $\to$ STU and then either LAT $\to$ CRC or LET $\to$ MLO. From STU to LAT early simulations showed that the virtual breast {\it does not} necessarily reach a symmetric configuration. This is because Evolver seeks to attain some pre-defined measurements and stabilizes right afterwards. The authors believe that the shape of the breasts naturally minimizes a certain energy whenever they rest at a given position. In the case of cell membranes, this energy is related to the {\it Willmore Functional} (for instance, see \cite{K}). This functional is defined as $\int_MH^2dS$, where $M$, $H$ and $dS$ stand for the surface of the body, its mean curvature and the area element of integration, respectively. By asking Evolver to minimize this integral we have obtained more symmetric answers. That is why the Willmore Functional is included in our datafiles. We conjecture that this energy is minimized whenever the breast accommodates in a rest position.

\subsection{Experiments with transparent breast phantoms}

As we have explained at the Introduction, this present work is devoted to simulations of external compressions of the breast. In order to track tumour nodules we shall either consider the complexity of internal parts, or equate their trajectory by performing experiments with transparent breast phantoms. 

Hence, instead of working with too many components we can take numerical trajectories and predefine them in our Evolver model. Figures 7 and 8 illustrate our strategy. 

\centerline{
\includegraphics[scale=0.45]{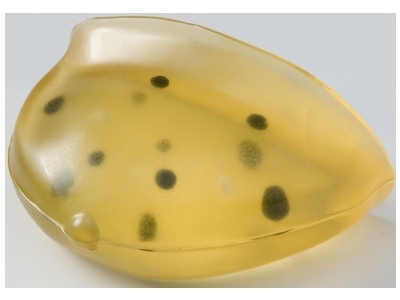}\hfill
\includegraphics[scale=0.37]{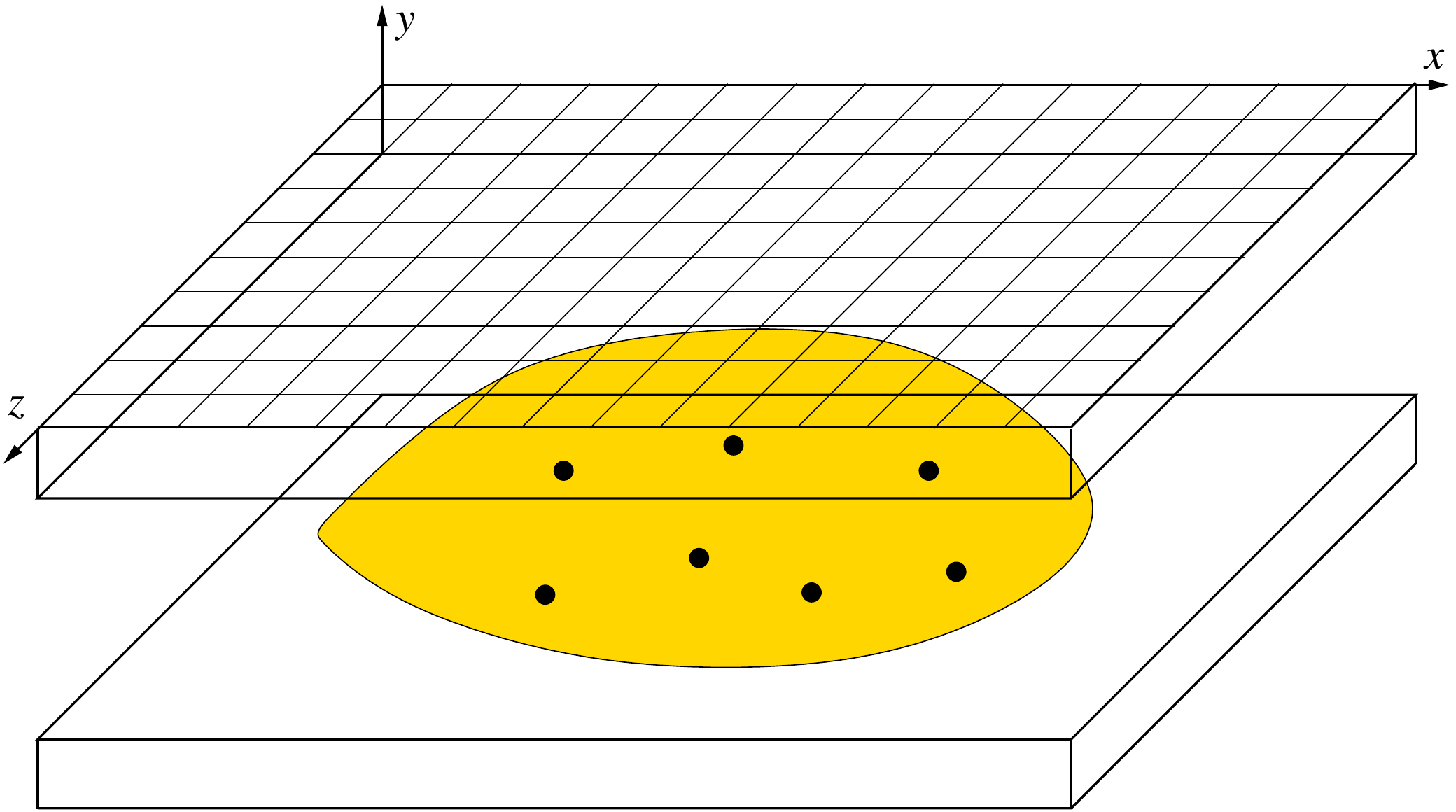}}

\centerline{Figure 7: A breast phantom\footnote{Stereotactic Needle Biopsy, taken from\\ {\tt http://www.imagingsol.net/product/1724/Stereotactic-Needle-Biopsy-Training-Phantom.html}}.\hfill Figure 8: Performing CC/MLO compressions.}
\ \smallskip

Compressions with the phantom were already performed and recorded in videos. We give more details in the next section.

\section{Results}

The main result is our programme, in which the user can track the first steps of a virtual mammography as depicted in Figure 9.
\\

\centerline{
\includegraphics[scale=0.30]{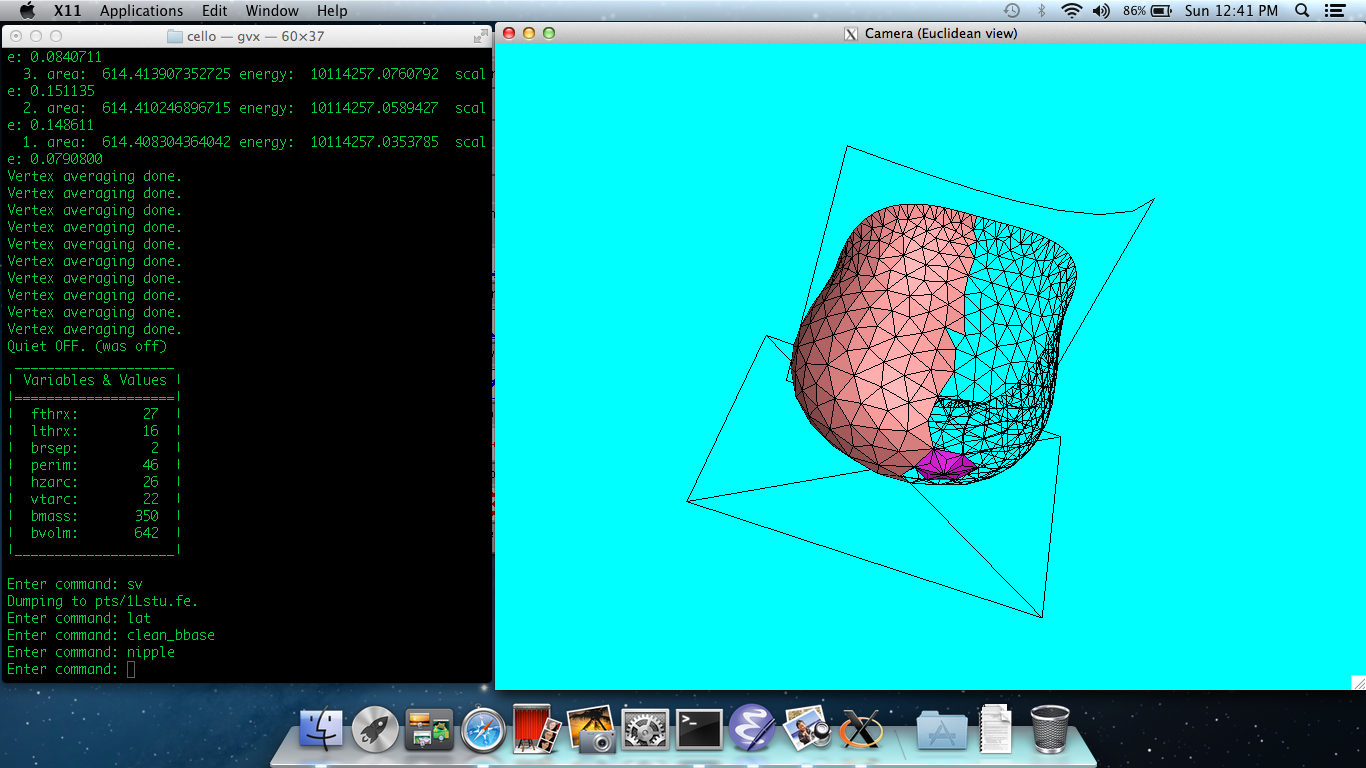}}

\centerline{Figure 9: LAT with displayed measurements.}
\ \smallskip

In order to run our programme we recommend Linux Operating System Ubuntu 11.10 or later, Evolver 2.70 and Geomview 1.9.4. For the time being, we have implemented SRG, STU and LAT, all of them for the left breast only. You can either run it in our Virtual Machine available on our home page 
\\

\centerline{{\tt http://www.facom.ufu.br/\~{}nascimento}}
\ \\
in the link {\it Softwares}, or access the link {\it Slides-ICMSQUARE} for an overview. The Virtual Machine contains all source codes and therefore all details of our modelling. 

Regarding its {\it validation}, the link Softwares also contains three sample videos of tests with a transparent phantom. The phantom dimensions are compatible with the volunteer's, and addition of virtual nodules to our model will be implemented in a forthcoming work.

It is important to notice that the breast volume {\it decreases} from SRG to STU, and again from STU to LAT. Indeed, when the woman is lain down on the operating table, the breast structure is uniformly distributed on its base. This is when we have the least pressure due to gravity.

The breast {\it mass} is constant and {\it mean density = mass/volume}. When the woman stands up, the breast mass concentrates at the bottom. Pressure due to gravity gets higher, and the same happens to the mean density. This makes the breast volume smaller. In the volunteer's case, our simulations with Evolver indicate that it decreases from 700cm$^3$ to 680cm$^3$ approximately.

Further on, when the woman lays her breast on the table, not only gravity by also the contact force acts. The volume decreases again. In the volunteer's case, it drops from 680cm$^3$ to nearly 660cm$^3$. Because of that, in this paper we have used the word {\it compression} as {\it the action of pressing something into a smaller space}, rather than {\it pressing or squeezing together}. This second meaning applies to our next paper, devoted to simulating the virtual CC and MLO mammographies. 

\section{Discussion}

Our model reproduces the same shapes and measurements really taken from the volunteer's breasts. Regarding the compressions with the phantom, special care must be taken with the trajectories of its nodules. The phantom is actually used in propaedeutics, and its uniform density is higher than the average density of the human breast. In fact, our experiments need phantoms that have elasticity, which is an important property of our anatomy. However, an artificial material satisfactorily close to the elasticity of the human breast is still unknown to us.

Figure 10 is a diagram that describes the main steps of our modelling. Basically, we intend to make the whole input consist only of values taken by a simple tape-measure, together with the mammographies. As the output, the model should give the precise location of nodules.
\\

\centerline{
\includegraphics[scale=0.90]{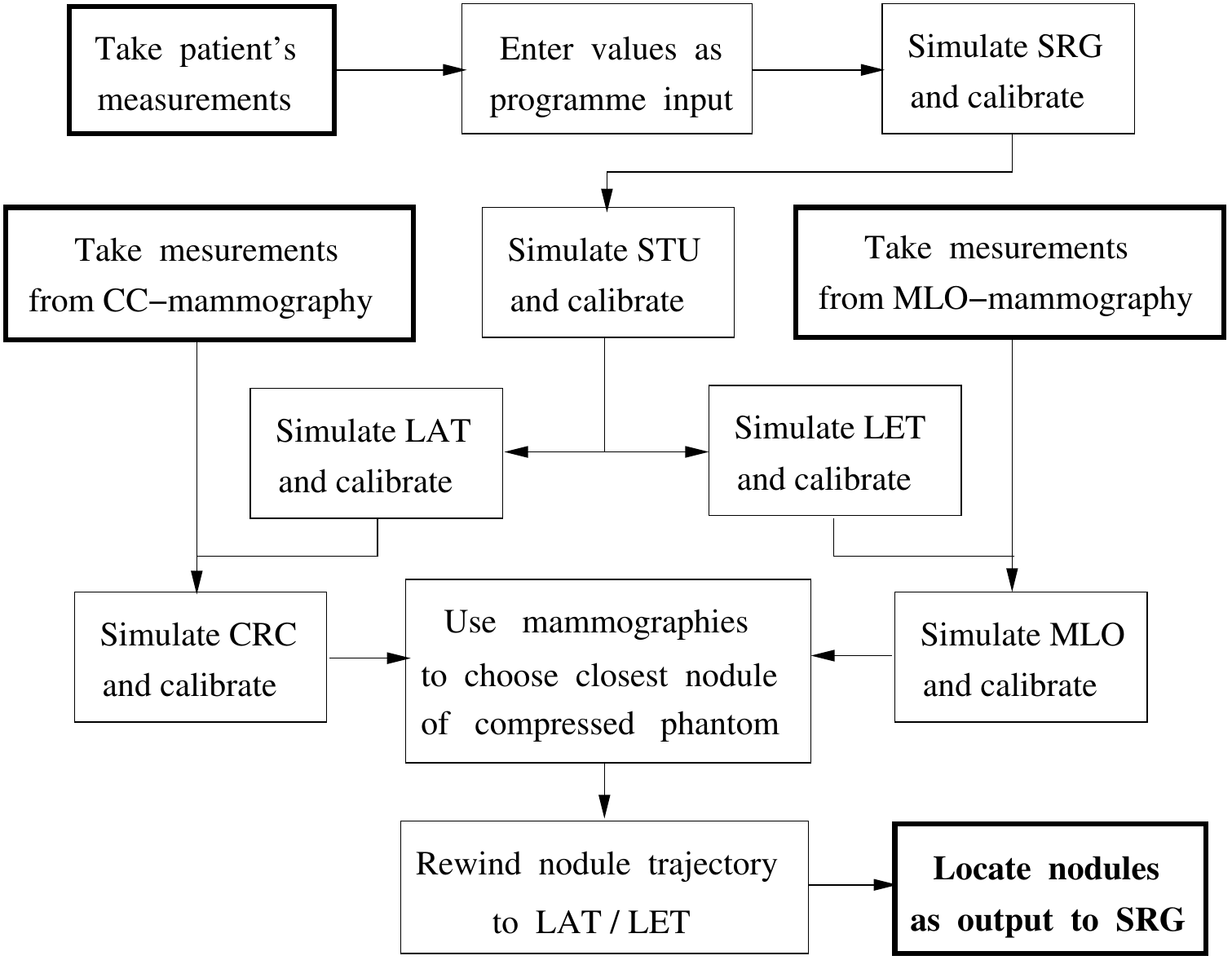}}

\centerline{Figure 10: Block scheme of the modelling.}
\ \smallskip

We have just presented the first results of a long-term research. These initial results are already of high complexity. For instance, nipple location, changes in shape, values and dimensions follow many geometric and physical laws. There are equated and implemented in the programmes, but technical details will be postponed to future publications.

\ack

We thank Dr Ana Cl\'audia Veronesi for details about gynaecological surgery, and her husband Alejandro Montepeloso for translating some technical terms. This research is supported by FAPESP proc.No.12/16244-3.

\section*{Appendix}

There is a virtual machine image called {\tt ubuntu11.10.ova} at
\\

\centerline{\tt http://hostel.ufabc.edu.br/\~{}marcelo.nascimento/software}
\ \\
for download. Get the programme {\tt VirtualBox} from Ubuntu Software Centre, start it and click on File/Import Appliance. Then pick up {\tt ubuntu11.10.ova} in the Download folder, open it, click on Next and then you should get the following window:
\\

\centerline{
\includegraphics[scale=0.30]{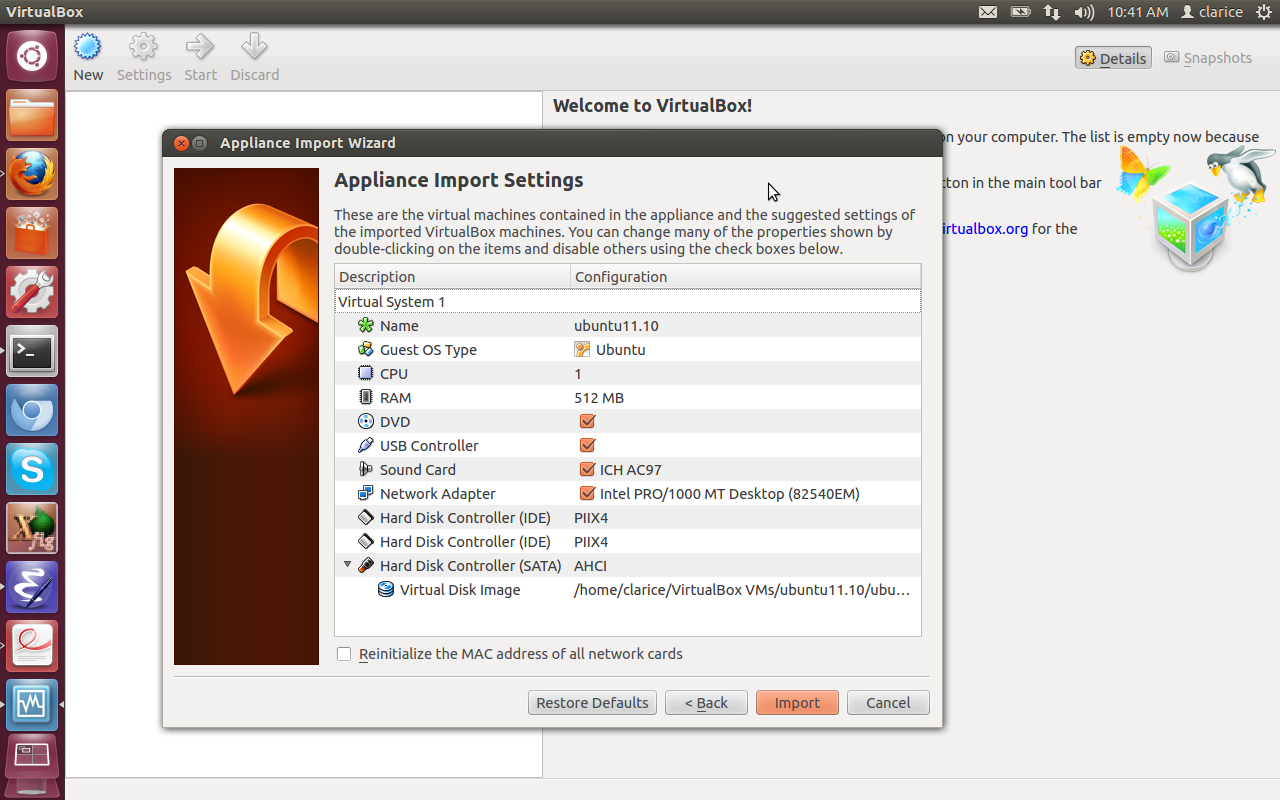}}
\smallskip

\centerline{Figure 11: Ready to import the virtual machine.}
\ \smallskip 

Click on Import and then VirtualBox will create {\tt ubuntu11.10}. However, when you start it there can be an error message like ``Implementation of the USB 2.0 controller not found!'' In this case, you can simply close the messages and go to the top left button of VirtualBox, named New. Give a name to your virtual machine, choose Linux, then the maximum amount of memory, and use your existing hard drive. Now click on ``Create'' and it will immediately appear on a list at the left-hand side. The list includes {\tt ubuntu11.10}, which you may remove later. Right after starting it, enter with the password ``evolver''. Once you get in the system, the shortcut Ctrl+Alt+T will open a terminal window already in the Evolver sub-directory. Our files are in the folder {\tt bsim}, a mnemonic for ``breast simulations''.

Of course, the virtual machine is never as fast as our personal computer itself. We are preparing a detailed manual with instructions for the readers who prefer to install the applications. The manual will be available on our homepage for download.

\end{document}